%Paper: 9201081
%From: LUGO@gaes.usc.es
%Date: Fri, 31 Jan 1992 20:21 GMT

%%%%%%%%%%%%%%%%%%%%%%%% NO MACROS ARE NEEDED !!! %%%%%%%%%%%%%%%%%%%%%%%%%

\magnification=1200
\font\text=cmr10
\font\it=cmti10
\font\title=cmbx10 scaled \magstep2

\font\caption=cmr8
\baselineskip=18pt
\text
\null \vskip 1cm

\def\ga{\alpha} \def\gb{\beta}  \def\gG{\Gamma} \def\gd{\delta}
\def\gD{\Delta} \def\ge{\epsilon}  
    \def\gm{\mu}
\def\gn{\nu} \def\gs{\sigma} \def\gS{\Sigma}  \def\gq{\theta}
  
\def\ex{{\hbox{\rm e}}}
 \def\plb{Phys. Lett. {\bf B}}
\def\n{\noindent}

\def\pMm{P_{\caption +}^{\gm\gn}}
\def\pmM{P_{\caption -}^{\gm\gn}}
\def\ac{\cal A}
\def\bc{\cal B}

\null
\rightline{ December 1991}
\bigskip\bigskip
\centerline {\bf COMMENTS ON GAUGED WZW MODELS}
\vskip 1.5cm
\centerline{ {A\caption DRIAN R. L\caption UGO}
\footnote*{M.E.C. of Spain fellow, \ \ E-mail: \caption LUGO@GAES.USC.ES} }
\bigskip
\centerline {\it Departamento de F\'\i sica de Part\'\i culas}
\centerline {\it Facultad de F\'\i sica, Universidad de Santiago}
\centerline {\it E-15706 Santiago de Compostela, Spain}
\vskip 2cm
{\caption
Some remarks are made about free anomaly groups in gauged WZW models.
Considerin
   g
a quite general action, anomaly cancellation is analyzed.
The possibility of gauging left and right sectors independently in some cases
is remarked. In particular Toda theories can be seen as such a  kind of
models.}

\vfill\eject

\n {\bf 1.} Among explicit realizations of conformal symmetry in two dimensions
[1], Wess-Zumino-Witten models (WZWM) [2] have been object of intensive study
in the past years, due to the fact they provide a realization of Kac-Moody (KM)
current algebra and the belief that any conformal field theory can be
represented by such a kind of theories [3].
Some gauged versions of them were considered, and it was shown that they lead
to
the Goddard-Kent-Olive coset models [4]. Recently an attempt to  classify all
"anomaly free" gauge subgroups was made in ref.[5]. In this  letter  we
generalize these results, and raise the possibility of considering  a wider
class of gaugings.
\bigskip

\n {\bf 2.} Let $G$ be a semisimple Lie group with Lie algebra $\cal{G}$,
$\gS$ a compact Riemann surface with metric
$\,\, h=h_{\gm\gn}d x^{\gm} \otimes dx^{\gn} \,\,$,  and g a mapping from
$\gS$ to $G$. WZW models are defined by the action [2]
$$ S_{\caption WZ}[g] \, = \, {k\over 8\pi} S_{0}[g] \, + \,
 i{k\over 12\pi}\gG[g]  \eqno(1a) $$
$$ S_{0}[g] = \int_{\gS} Tr(U(g) \wedge {*}U(g)) \eqno(1b) $$
$$ \gG[g] = \int_{B} Tr(U(g)\wedge U(g)\wedge U(g)) \ \ \ , \  \partial B=\gS
\eqno(1c) $$
where $"{*}"$ stands for the Hodge mapping wrt $\, h$ [6], $\, U(g)=g^{-1}dg$,
$ {\, \tilde U}(g)=dg g^{-1}$, and "Tr" for a conveniently normalized scalar
product on $\,\cal G$. The equations of motion for $U(g)$ that follow from
$(1)$
are $\, (d{*}+id)U(g)=0 \,$, and yield the usual conservation of the currents
$$     {\cal J}^{\gm} \equiv  {k\over 8\pi} {\pmM} {\tilde U}_{\gn}(g),
\, \, \, \, \, \,  {\bar {\cal J}}^{\gm} \equiv  {-k\over 8\pi} {\pMm}
U_{\gn}(g) \eqno(2a) $$
$$ \nabla_{\gm} {\cal J}^{\gm} \, = \,
g^{-1} \, {\nabla}_{\gm} {\bar{\cal J}}^{\gm} \, g \,=\, 0 \eqno(2b) $$
where $\, P_{\pm}^{\gm\gn}\equiv h^{\gm\gn} \pm i{\ge}^{\gm\gn}/{\sqrt h}\,$.
\bigskip
\n {\bf 3.} The action (1) is invariant under global transformations:\ \
$ g \rightarrow h_{L}g h_{R}$, with \ \ $h_{L} (h_{R})$ belonging to any
subgroup $H_{L} (H_{R})$ of $G$. Weyl invariance of the theory extends these
transformations to  holomorphic (anti) holomorphic dependence of $h_L$($h_R$),
being the generators of these transformations the momenta of $\cal J$
(${\bar{\cal J}}$), that satisfy the standard  level $k$ left (right) KM
algebra.
\footnote{$^1$}{The reader notes that in isothermal coordinates the only
non vanishing  components of $\, P\,$ are
$\,\, P_{\caption +}^{z \bar z} = P_{\caption -}^{{\bar z} z} =
  2 {h}^{z\bar z}$,  so that (2b) asserts the holomorphicity (anti) character
of ${\cal J}$ (${\bar{\cal J}}$).}

 Now we would like to write an action invariant under {\it arbitrary\/}
$h_{L},\, h_{R}$.
Having two conserved currents, the natural thing to do for gauging the theory
is to couple them to gauge fields $\ac$ and $\bc$ valued in the Lie
subalgebras ${\cal H}_R$ and ${\cal H}_L$ respectively, and consider the
gauge transformations
$$ g^{(h)} = h_L\  g\  h_R  \eqno(3a) $$
$$ {\ac}^{h_{R}} = {h_{R}}^{-1} \, {\ac}\, h_{R} + U( h_{R} ) \eqno(3b) $$
$$ {\bc}^{h_{L}} = h_{L} \, {\bc} \, {h_{L}}^{-1} + {\tilde  U} ({h_{L}} )
\eqno(3c) $$

Let us now take as the gauge action (dictated by locality and invariance
arguments and, of course, previous knowledge on  the subject)
$$ {S_{\caption G}}[g,{\ac} ,{\bc} ] = {k\over 4\pi} \int_{\gS} d^{2} x
\sqrt{h}
\,\, Tr[\, - \pMm {\ac}_{\gm} U_{\gn} (g) - \pmM {\bc}_{\gm} {\tilde U}_{\gn}
(g
   )
\,+$$
$$ {\pMm} g {\ac}_{\gm} {g}^{-1} {\bc}_{\gn} \, +\,
{1\over 2} h^{\gm\gn}  ({\ac}_{\gm}  {\ac}_{\gn}
+ {\bc}_{\gm} {\bc}_{\gn}) \, +\, h^{\gm\gn} (\, {\ac}_{\gm} {\cal R}_{\gn}\,
-\
   ,
{\bc}_{\gm} {\cal S}_{\gn}\, )\,] \eqno(4) $$
where $\cal R$ and $\cal S$ are (in principle) arbitrary ${\cal G}$ valued
1-for
   ms.

By using the Polyakov-Wiegmann formula
$$ S_{\caption WZ}[gh] = S_{\caption WZ}[g] +  S_{\caption WZ}[h] +
{k\over 4\pi} \int_{\gS} d^{2}x \sqrt{h}\,  Tr[ {\pmM}\, U_{\gm}(g) \,
{\tilde U}_{\gn}(h)] \eqno(5) $$
is straightforward to get the variation of \ \
$ S = S_{\caption WZ} + S_{\caption G}\, $ under (3). We get
$$ S[g^{(h)}, {\ac}^{h_{R} }, {\bc}^{h_{L}}] - S[g,{\ac},{\bc}]\,
= \,  i{k\over 12\pi} (\gG[ h_R ] +\gG [ h_L ] ) \, + $$
$$+ \,  i{k\over 4\pi} \int_{\gS} Tr[ {\bc} \wedge U( h_{L} ) -
{\ac} \wedge {\tilde U} (h_{R})] \, + $$
$$ +\,{k\over 4\pi} \int_{\gS} d^{2}x \sqrt{h}\,  h^{\gm\gn}\,\,
Tr[\, ( {\ac}_{\gm}^{h_{R}}  -
{\ac}_{\gm} ) {\cal R}_{\gn} \, - \, ({\bc}_{\gm}^{h_{L}} - {\bc}_{\gm} )
{\cal S}_{\gn} ) ] \eqno(6) $$

\n The RHS of (6) constitutes the "anomaly". As pointed out in [5], the first
term being non local should vanish by itself.
Up to direct products, there are two way of getting this:
\bigskip
\n${\underline a)}$ $\, \gG[h_R ] = -\gG[h_L ] \ne 0\,$.

This yields $\, h_R =h_L^{-1}\equiv h\, $, and so we must take also
$\bc =-\ac\,$ (and shift ${\cal R}\rightarrow {\cal R}-{\cal S})\,$;
the anomaly becomes
$$ RHS(6) = {k\over 4\pi} \int_{\gS} d^{2}x\sqrt{h}\,
Tr[\, h^{\gm\gn} ( {\ac}_{\gm}^{h} - {\ac}_{\gm} ) {\cal R}_{\gn} ]\, \eqno(7)
$
   $

The choice $\,{\cal R} = 0 \,$ is well known in the literature, the so called
"vector" case. The equations of motion derived from (1,4) in the gauge
$\, {\cal A} =0 \, $
\footnote{$^2$}{Here and in what follows we will consider this kind of gauge,
having in mind  that only on trivial topologies is possible to reach it; at
higher genus the "harmonic" sector (related to non contractible loops on $\gS$)
can not be gauged away.}
are (2b) supplemented by the constraint equation
$$ \, J(T_a ) + {\bar J}(T_a ) = 0\, \eqno(8) $$
where $\,  J(T)\equiv Tr({\cal J}T)\,$ and $\{ T_a \}$ are the generators of
$\cal H\,$.

Other choices are also possible, for example, if ${\cal H}$ is abelian,
${\cal R}\,$ can be taken as ${\cal R} = {\cal E} + {\bar {\cal E}}\, $ with
${\cal E}$ holomorphic and $\, {\bar {\cal E}} \, $ its conjugate. The
constraint equations now read
$$ J(T_a ) \, -\, {\bar J}(T_a) \, = {k\over 8\pi} Tr(\, T_a {\cal R}\, )
\eqno(9) $$
Another one, to be considered below in the case b), is to take
$\,{\cal H}\, $ nilpotent.
\bigskip
\n ${\underline b) } $ $ \, {\gG} ( h_{R} ) \, = \, {\gG} ( h_{L} ) = 0 \, $.

One way to get it is to consider abelian subgroups
$\,{\cal H}_R = {\cal H}_L \equiv {\cal H}$; for
$\, h_L = {h_R}^{-1} \equiv h\, $ and $\,{\cal B} = -{\cal A}\, $
we are in the vector case, but also we can choose
$\, h_L = {h_R} \equiv h\, $ and $\,{\cal B} = {\cal A}\, $,
the so called axial case; the conditions to cancel the anomaly are the same
as before, and the constraints equations also looks like (9); both theories are
dual [5].

Another interesting possibility  is to consider $\, {\cal H}_L \,$ and $\,
{\cal
    H}_R
\,$ to be  nilpotent; in this case the scalar product among any of their
element
   s
vanish  (see [7], page 13). The anomaly (6) reduces to the last term.
\footnote{$^3$}
{Note also that the cuadratic terms in the gauge fields in the action (4) go
down.}
It seems there is no way to reach its cancellation in general, but
there is a case in which it ocurrs. Let us consider the group $SL(N,R)$, the
generators of its Lie algebra $\, {\cal G} = {\cal L}_{-} + {\cal C} + {\cal
L}_
   {+}
\,\,$, $ \, {\cal L}_{-} = \{ \, E_{-\ga}, \ga \in {\Phi}^{+}\}, \, \,
  {\cal L}_{+} = \{ \,  E_{\ga}, \ga \in {\Phi}^{+}\}, \, \,
      {\cal C} = \{ H_{\ga}, \ga\in \gD \} $
where ${\Phi}^{+} (\gD)$ stands for the set of positive (simple) roots,
and choose $ \, {\cal H}_R = {\cal L}^+ , \, {\cal H}_L = {\cal L}^- \,$,
being $\, {\cal L}_{\pm}\,$ nilpotent (see [7], page 37).
If we take ${\cal R}\in {\cal L}_-, \, {\cal S}\in {\cal L}_+ \,$ the
constraint

equations read
\footnote{$^4$} { In what follows:
 $ \, h_{R} = \ex^{  {\ge}_{R}^{\ga}  E_{\ga}  }, \,
  h_{L} = \ex^{  {\ge}_{L}^{\ga} E_{-\ga}  }  \, \, \, ;
{\cal R} = R^{\ga} /a \, E_{-\ga} \, \, \, , \, {\cal S} = S^{\ga}/a E_{\ga}
\,\
   , ;
Tr( E_{\ga} E_{-\gb} ) = a \gd_{\ga,\gb} \,\,$; all the roots and implicit sums
in $\,\Phi^+ \,$.}
$$ {\cal J}( E_{\ga} ) = - {k\over 8\pi} R_{\ga} \,\,\,\,\,
   {\cal J}( E_{-\ga}) =   {k\over 8\pi} S_{\ga} \eqno(10) $$

In ref.[8] was shown that by making the special choice: $\, S_{\ga} , \,
R_{\ga}
    \, $
arbitrary constants non zero for $\ga \in \gD$ and zero otherwise, eq.(2b)
supplemented by (10) are equivalent to the Toda field equations (in the
variable
   s of
the Cartan subalgebra ${\cal C}$). We have verified for $N=2,3$ (and conjecture
   for
any N) that when these conditions are verified, the anomaly takes the form

$$ RHS(6) = {k\over 4\pi} \int_{\gS} d^{2}x \sqrt{h} \,
Tr[\, h^{\gm\gn} ( {\ge}_{L}^\ga {\nabla}_\gm {\cal S}_{\gm}^\ga \, - \,
{\ge}_{R}^\ga {\nabla}_\gm  {\cal R}_{\gn}^\ga )\,] \eqno(11) $$
from where we get the cancellation for general holomorphic (anti)
${\cal R}({\cal S})$ (in order to be compatible with the constraints (10)).
This result can be seen as a  generalization, including the higher genus case.
\bigskip
 As an example, let us take the simplest one, $G=SL(2,R)$.
In the (local) parametrization
($\gs_\pm = \gs_1 \pm \gs_2 \,; \gs_i $ being the Pauli matrices)
$$ g = \ex^{ \gq_L \gs_{-} } \ex^{ r\gs_{3} } \ex^{ \gq_R \gs_{+} }\, , \,\,
    {\cal A} = A \gs_{+} \,\, , \,\,{\cal B} = B \gs_{-}    \eqno(12) $$
the action takes the form
$$ S \, = \, {k\over 4\pi} \int_{\gS} d^{2}x \sqrt{h} \,
[ \, {\partial}_{\gm} r {\partial}^{\gm} r \, + \, \ex^{2r} \,
\pMm  (A_{\gm} - \partial_{\gm} \gq_R)  (B_{\gn} - \partial_{\gn} \gq_L )
+  h^{\gm\gn}  (A_\gm R_\gn - B_\gm S_\gn )  ] \eqno(13) $$
Integrating out the gauge field it is not dificult to see that the effective
act
   ion
in the $\, r$ variable defined by
$$ \ex^{-W[r]} \, = \, \int {\cal D}{\cal A} {\cal D}{\cal B}{\cal D}\gq_R
{\cal D}\gq_L \,\, \ex^{-S} \eqno(14) $$
has the form of the Liouville action, with cosmological
constant proportional to the product of the non zero components of ${\cal R}$
an
   d
${\cal S}$, and a shift of $\, k\rightarrow (k-2)$ in the kinetic term (see [9]
   for
details to this respect).
However some subtelties are present here. First, in the euclidean language
we are working, we must assume that the integrals of linear exponentials lead
to
delta functions; second, we must divide out by the gauge group, also taking
into
account that only one component of ${\cal A}$ and ${\cal B}$ are present in
(14)
   .
Instead of following this procedure (that however gives an insight of what the
effective target space is) to quantize these kind of theories, we could
adopt that, in somewhat safer, pursued in ref.[10] to study the black hole
solut
   ion
found in [11]; analysis of the spectra seems to be reachable (and in fact, it
wa
   s
carried  out also in the above example, considering a gauged WZWM slightly
diffe
   rent of
ours and studying their primary fields), but the amplitudes involved to
compute.

\bigskip
\n{\bf Acknowledgements.} We thank valuable and enlightning conversations with
J. Labastida, J. Mas and A. V. Ramallo.

\vfill\eject

\n{\bf References}

\n[1] For a review, see: L. Alvarez-Gaume, C. Gomez and G. Sierra,
"Topics in Conformal Field Theory", preprint CERN-TH.5540/89.

\n[2] E. Witten, Comm.Math.Phys. {\bf 92} (1984), 455.

\n[3] For a review, see: P. Goddard and D. Olive, Int.Jour.Mod.Phys. {\bf A}1
      (1986), 303.

\n[4] D. Karabali: "Gauged WZW models and the coset construction of CFT",
Brandeis preprint (July 1989) and references therein.

\n[5] E. Kiritsis: "Duality in gauged WZW models", preprint LBL-30747, May
1991.

\n[6] T. Eguchi, P. B. Gilkey and A. J. Hanson, Phys. Rep. {\bf 66}
      (1980), 213, and references therein.

\n[7] A. Barut and R. Raczka: "Theory of group representations and
applications", World Scientific (1986).

\n[8] P.Forg\'acs et al., \plb 227 (1989), 214.

\n[9] A. Gerasimov et al., Int.Jour.Mod.Phys. {\bf A}5 (1990), 2495.

\n[10] R. Dijkgraaf, H. Verlinde and H. Verlinde: "String propagation in a
black
hole geometry", IAS preprint (March 1991).

\n[11] E. Witten: "On string theory and black holes", IAS preprint (March
1991).

\bye\supereject\end